\documentstyle[aps,prb,floats]{revtex}
\begin{document}
\renewcommand{\textfraction}{0.10}
\renewcommand{\topfraction}{1.0}
\renewcommand{\bottomfraction}{1.0}
\flushbottom

\twocolumn[
\title{Holons in chiral spin liquids: statistics and pairing}
\author{D.M. Deaven, D.S. Rokhsar}
\address{Department of Physics, University of California, 
Berkeley, CA 94720}
\author{A. Barbieri}
\address{Lawrence Berkeley Laboratory, Berkeley, CA 94720}
\date{\today}
\maketitle
\begin{abstract}
\widetext
\advance\leftskip by 57pt
\advance\rightskip by 57pt

We study Gutzwiller-projected variational wavefunctions for charged, 
spinless holon excitations in chiral spin liquids. We find that these
states describe anyons, with a statistical phase $\Phi_s$ that is 
continuously adjustable between $0$ and $\pi/2$, depending on a 
variational parameter.
The statistical flux attached to each holon is localized to within a 
lattice constant.
By diagonalizing the effective Hamiltonian for charge motion in our
variational basis we obtain a two-holon pair state.
This two-holon state has large overlap with a two-hole state 
whose relative wavefunction has $d$-wave symmetry and breaks 
time-reversal and parity.
\end{abstract}
]

\narrowtext

\section{INTRODUCTION}
\label{sec_intro}

The notion of a quasiparticle is a remarkably successful
organizing principle for understanding many-body systems.
In strongly interacting Fermi systems such as $^3{\rm He}$, for
example,
many-body effects simply renormalize the mass and interactions of the 
Landau quasiparticles;  their charge, spin, and statistics the same as
those of the bare atoms.\cite{Landau}
Other many-body systems, however, possess low-energy charged
excitations
that are dramatically different from Landau quasiparticles.
In the fractional quantum Hall effect, for example, a large
magnetic field 
and the Coulomb repulsion between electrons conspire to produce
Laughlin
quasiparticles with fractional charge and statistics.\cite{FQHE}
In {\it trans-}polyacetylene, a system with broken translational
symmetry, 
charges bind to domain walls, forming soliton excitations with charge 
$e$ but without spin.\cite{ssh}
In both of these cases, the novel quasiparticles have a
``topological''
character, because they cannot be created by local combinations of 
bare electron creation and destruction operators, in contrast with
Landau quasiparticles. Nevertheless,
these unconventional quasiparticles
are well defined, weakly interacting objects which 
describe the low-energy states of the many-body system.

Similarly, unconventional quasiparticles have been suggested as the 
charge carriers in ``resonating valence bond'' theories of high 
temperature superconductivity.\cite{rvb,krs,kl} 
Motivated by the fact that cuprate high temperature superconductivity 
arises upon doping antiferromagnets exhibiting large quantum
fluctuations, 
these theories focus on spin-$1/2$ systems with local
antiferromagnetic 
correlations but no long range order -- so-called ``spin liquids.''  

The phase diagram of the high temperature superconductors motivates 
the study of spin liquids as a stepping stone in a novel
approach to the
superconducting state. Referring to fig. \ref{scenario},
the real materials reside in the 
temperature-doping plane; an additional axis, labeled frustration,
is added for theoretical convenience.
When the parent insulators are doped, antiferromagnetic long range
order is destroyed almost simultaneously with the onset of 
superconducting order.  (A ``spin glass'' regime due to disorder
may intervene.)  It is appealing to try to separate the effects of
loss of magnetic order from the instability to superconductivity.
In principle, this could be accomplished theoretically by first
introducing fictitious frustrating interactions in the
antiferromagnetic
insulator, thereby destroying long range order without
introducing charge
carriers.  Then, to study the appearance of superconductivity, this
``spin liquid'' state would be doped, so that one could consider the
motion of dilute charge carriers in a spin background resembling
the fluctuating spins in the real materials' superconducting state.

Finally, to return the theory to a more realistic model of the actual
copper oxide planes, the fictitious frustrating interactions would be 
turned off -- they would be no longer needed, since magnetic
frustration
is implicitly generated by charge carrier motion.\cite{inui}
In this roundabout manner, one would arrive at a description of the
real materials' superconducting state.  Assuming that no phase
transition intervenes, the superconducting state achieved upon doping
the spin liquid would have the same symmetry and general features as
the real superconducting state.  This approach contrasts the
more direct
``BCS approach,'' which begins from an understanding of the
normal metal
at high temperature (which may be problematic in the cuprates) and
then considers residual attractive interactions between
quasiparticles.
If properly executed, both the spin liquid and the BCS approaches
should
arrive at the same superconducting state, but from different
directions.

Unfortunately for the spin liquid approach, it has proven difficult to
find suitably frustrated spin models which destroy antiferromagnetic
order without introducing additional unwanted broken symmetry ground
states.  There is no short-ranged spin model known to have
a spin liquid
ground state.  Nevertheless, spin liquids may be usefully thought
of as 
metastable phases of frustrated antiferromagnets that mimic the spin
correlations of cuprate superconductors, even if they are not
the true ground state of a particular
frustrated spin system.  One can still study 
their excitation spectra,
and hope to carry out the program outlined above.  In this work we
consider simple variational ``chiral spin liquid'' 
states\cite{wwz} which break
time-reversal and parity symmetries but are translationally invariant.
They have good variational energies for simple frustrated
spin Hamiltonians,
and suggest natural variational states when doped.  By studying the 
motion of a few charge carriers, we develop a scenario
for understanding
superconductors with broken time-reversal and parity symmetry.

\begin{figure}
\setlength{\unitlength}{0.0125in}%
\begin{picture}(260,235)
\end{picture}
\caption[scene]{
Schematic phase diagram for high temperature superconductivity.
The real materials occupy the temperature-doping plane; an additional
axis, labeled frustration, is included for theoretical convenience.
In a BCS-like approach, the superconducting state is entered from a
well understood normal state.  In the ``resonating-valence-bond''
or ``spin liquid'' approach, a disordered quantum antiferromagnet
is stabilized by artificially adding magnetic frustration to the
undoped parent compound.  This spin liquid is then doped to form a
superconductor.  The condensate is formed from bound holon pairs, 
the charge carrying quasiparticles in a doped spin-liquid. If no
phase transition intervenes as the frustration is turned off in the
doped state, both the spin liquid and BCS approaches must arrive
at the same superconducting state, but from complementary physical 
points of view.}
\label{scenario}
\end{figure}

We consider the $t$-$J$-$J'$ Hamiltonian 
${\cal H}={\cal H}_{\rm spin} + {\cal H}_{\rm hop}$, where
the first term is the (frustrated) spin Hamiltonian
\begin{equation}
{\cal H}_{\rm spin} = \sum_{\langle ij\rangle} J_{ij}
{\bf S}_i\cdot {\bf S}_j ,
\label{htj1}
\end{equation}
with first and second neighbor anti\-ferromagnetic couplings $J$ 
and $J'$, and the second term is the
nearest-neighbor hopping Hamiltonian
\begin{equation}
{\cal H}_{\rm hop} = -t\sum_{\langle ij\rangle, \sigma} 
c^\dagger_{i\sigma} c_{j\sigma}.
\label{htj2}
\end{equation}
In these expressions $c^\dagger_{i\sigma}$ creates a
fermion at site $i$ 
with spin projection $\sigma$ and
${\bf S}_i$ $\equiv$ $c^\dagger_{i\alpha}
\sigma_{\alpha\beta} c_{i\beta}$
is the spin operator at site $i$.  Both Eqs.\ (\ref{htj1}) 
and (\ref{htj2})
are allowed to act only within the subspace of states in which no
lattice site is doubly occupied.

In this work we address
(a) the internal structure of charged ``holon'' excitations
in a locally stable spin-liquid and (b) the
dynamics and statistics of a dilute gas of
these excitations.  Sections \ref{sec_general} and \ref{sec_holons}
contain a brief general discussion of spin liquids and holon
excitations and describe the specific wavefunctions we consider.
Our main results are presented in sections \ref{sec_statistics} and
\ref{sec_pair}.  A brief summary is provided in
section \ref{sec_conclude}.

\section{CHIRAL SPIN LIQUIDS}
\label{sec_general}

Spin liquids are quantum antiferromagnets whose N\'eel order has been
destroyed by large zero-point fluctuations of the interacting spins.  
These large quantum effects preclude the use of a spin-wave expansion 
about an ordered state.  Instead, we appeal to 
the variational principle and take advantage of qualitative
similarities between spin liquid states and certain simple
Slater determinant
states of spin-$1/2$ fermions.  In particular, consider
Slater determinant states $|\chi\rangle$ which 
(a) are translationally invariant with
an average density of one fermion per site, (b) have no average
magnetic moment at each site, and (c) have spin and
charge correlations
which decay exponentially. All three of these properties are
shared by 
spin liquids.  The main difference between $|\chi\rangle$ and
a spin liquid is simply one of degree -- a spin liquid has no number 
fluctuations at all, while $|\chi\rangle$ has local
number fluctuations.

This distinguishing property of $|\chi\rangle$ can be eliminated by 
applying the complete Gutzwiller projector
\begin{equation}
P_{\rm G} \equiv \Pi_i (1 - n_{i\uparrow} n_{i\downarrow}),
\label{gutzpro}
\end{equation}
which annihilates all configurations containing doubly-occupied sites.
What remains is a superposition of spin configurations whose 
amplitudes are inherited from the original ``pre-projected''
Slater determinant.
If this pre-projected state has short-ranged spin and
charge correlations, 
then Gutzwiller projection is a ``local'' procedure.\cite{localgutz}
The correlations in the projected state then closely follow
the correlations 
of the pre-projected determinant, up to simple renormalization 
factors of order unity.\cite{gutzwiller,other-cluster,us}

\begin{figure}[t]
\setlength{\unitlength}{0.0125in}%
\begin{picture}(160,405)
\end{picture}
\caption{
Schematic of a tight binding model $H_\chi$
which breaks time-reversal and parity
symmetries. Circles represent sites; lines represent hopping
matrix elements between sites. A line with an arrow has phase $\pi/4$
in the sense given by the arrow, a dashed line has phase $\pi$, and a
solid line zero phase. (a) The flux state. Each elementary plaquette
has $\pi$ flux piercing it. (b) The chiral state. Each elementary
triangle has flux $\pi/2$ piercing it.
Up to a gauge transformation, these tight-binding Hamiltonians are
translationally invariant.}
\label{fluxtbm}
\end{figure}

It is surprisingly difficult to construct Slater determinants with 
the three properties listed above. 
To obtain a translationally invariant free-particle state with 
only short-range correlations, one should completely fill a band of 
single-particle states while simultaneously
ensuring that a gap exists between the filled 
band and the next highest energy band.
For a translationally invariant system, the resulting charge density
will be uniform.  
The difficulty arises in constructing a translationally
invariant system with a one-body energy gap.
The simplest way to accomplish this is to 
consider particles moving
on a lattice in a uniform commensurate (fictitious)
magnetic field which 
couples only to orbital motion.\cite{wwz,marston}
For appropriate field strength, the
effective unit cell is doubled, and a gap is opened at half-filling.
The hopping matrix elements will now be complex;
as long as the net phase factor in the product of the 
elements around any closed loop is invariant under
translations, the model ground state will also be translationally
invariant, up to an overall gauge transformation.  

In particular, let $|\chi\rangle$ be the half-filled ground state of
the square-lattice tight-binding Hamiltonian\cite{wwz}
\begin{equation}
H_\chi = \sum_{ij\sigma}\chi_{ij} c_{i\sigma}^\dagger c_{j\sigma}
\label{hmf}
\end{equation}
illustrated in fig. \ref{fluxtbm}.
Here $\chi_{ij}$ is a complex link variable and
$c^\dagger_{i\sigma}$
creates a fermion at site $i$ with spin projection $\sigma$.
We arrange for a net phase $\pi$ around every plaquette, and phase
$\pi/2$ around every elementary triangle (encircled clockwise, say). 
More explicitly, the complex product
$\chi_{AB}\chi_{BC}\chi_{CD}\chi_{DA}$ 
around any elementary plaquette $ABCD$ has phase $\pi$, 
while the product
$\chi_{AB}\chi_{BC}\chi_{CA}$ around any elementary triangle $ABC$
has phase $\pi/2$.
We will refer to the phase around a closed loop
a ``flux,'' since the (Aharonov-Bohm) phase for motion in a 
magnetic field is proportional to the flux enclosed by the path.
One flux quantum $hc/e$ corresponds to a phase $2\pi$.

The Hamiltonian Eq.\ (\ref{hmf}) explicitly breaks both time-reversal 
and parity symmetries, each of which inverts the flux through every 
closed loop.  
In this work we shall set $\chi_{ij}=0$ for all but nearest
and next-nearest neighboring sites $i$ and $j$.  We set 
$|\chi_{ij}|=1$ when sites $i$ and $j$ are nearest neighbors, and
$|\chi_{ij}|=m_0/2$ when sites $i$ and $j$ are next-nearest
neighbors.
This notation derives from the fact that for small values of $m_0$
and at half filling,
the noninteracting fermions whose motion is governed by
Eq.\ (\ref{hmf})
have a gap to excitations equal to $4m_0$ (``mass,'' in
field-theory jargon).
Figure \ref{chiralband} illustrates the gap opening due to 
the addition of the diagonal hopping strength $m_0$.

\begin{figure}
\setlength{\unitlength}{1mm}%
\begin{picture}(70,73)
\end{picture}
\caption[fig_band-cap]{
Single-particle bands of Eq.\ (\ref{hmf}) in the two-site
Square $\sqrt 2\times\sqrt 2$ magnetic unit cell
shown by the dotted line in the inset, with no background magnetic
field (dashed line) and in the presence of a translationally invariant
chiral field (solid line) as illustrated in
fig. \ref{fluxtbm}. Note that a gap equal to $4m_0$ is opened at
half-filling (filling the lower band).}
\label{chiralband}
\end{figure}

The similarity of the Gutzwiller-projected states $P_G|\chi\rangle$
to the actual ground states of a doped antiferromagnet may be measured
by their variational energy.
Consider ${\cal H}$ with $J_{ij}=J$ for nearest neighbor sites
$i$ and $j$, and $J_{ik}=J'$ for next-nearest neighbor sites
$i$ and $k$. Then
for each value of $J'/J$ there is a corresponding
value of $m_0$ which minimizes the variational energy
\begin{equation}
E(J'/J ; m_0)=
\left.{\langle\chi|P_G{\cal H}_{\rm spin}(J'/J)P_G|\chi\rangle \over
\langle\chi|P_G|\chi\rangle}\right\vert_{m_0}.
\label{evary}
\end{equation}
In this paper we study systems
with values of $m_0$ satisfying $0<m_0\le 1$,
which implicitly selects the value of $J'/J$ via
Eq.\ (\ref{evary}) (see ref.\onlinecite{us} for a more complete
description of the properties of $E(J'/J ; m_0)$). For small $J'$ 
it is generally believed\cite{nosl} that the ground state of 
${\cal H}_{\rm spin}$ 
has N\'eel order, unlike the variational states we consider here.  
For larger $J'$ other more complicated types of broken symmetry
ground states may be stabilized.
For our purposes, we assume that the spin liquid states 
$P_G|\chi\rangle$
are adequate models for the spin background of a lightly doped quantum
antiferromagnet.

The parameter $m_0$ plays a crucial role in our calculation.
It corresponds to the gap in the spectrum of $H_\chi$ at
half-filling (see fig. \ref{chiralband}) and determines the
decay length $\xi_\chi$ for exponentially decaying spin and charge 
correlations in the pre-projected state $|\chi\rangle$.
For small $m_0$, $\xi_\chi\approx 1/m_0$.
(For general values of $m_0$, $\xi_\chi(m_0)$ has a more
complicated but nonincreasing form\cite{us} when $0<m_0\le 1$.) 
The spin-spin correlation length in the {\it projected} state 
$P_G|\chi\rangle$ is related to $\xi_\chi$ by a numerical factor of 
order unity,\cite{us} so $m_0$ also adjusts the
range of spin correlations in the fully projected state.

The introduction of a non-zero $m_0$ in Eq.\ (\ref{hmf}) breaks both
T (time-reversal) and P (two-dimensional parity) symmetry.\cite{wwz}
This can be seen, for example, by computing the spin triple product
$\langle {\bf S}_i \cdot {\bf S}_j\times{\bf S}_k\rangle$
on any elementary triangle $ijk$ in the projected state.  
This expectation value vanishes by symmetry when $m_0$ vanishes,
but is non-zero otherwise.

\section{SPINONS AND HOLONS}
\label{sec_holons}

Unlike the creation of quasiparticles in a Fermi liquid, the
creation of separated spin and charge excitations in spin liquids
involves non-local operations.
There are two schemes for producing holons and spinons 
in Gutzwiller projected states, which at first glance
seem quite different.

{\it Scheme 1} -- Anderson\cite{anderson} begins with 
a locally non-neutral state of the form
\begin{equation}
c_{i\,-\sigma} c_{j\sigma'}^\dagger |\chi \rangle,
\label{prespinon}
\end{equation}
a Slater determinant to which a particle
with spin projection $\sigma'$ has been added at site $j$ and
a particle with spin projection $-\sigma$ has been removed
from site $i$. In this state,
$\langle n_j\rangle=3/2$ and $\langle n_i\rangle=1/2$, while $\langle
n_k\rangle=1$ for all other sites $k$, as in the original Slater
determinant state $|\chi\rangle$.  The projected state
\begin{equation}
|i\sigma ,j\sigma' \rangle \equiv P_{\rm G} 
c_{i\,-\sigma} c_{j\sigma'}^\dagger |\chi \rangle
\label{spinon}
\end{equation}
has a uniform charge density everywhere.  The act of projecting
Eq.\ (\ref{prespinon}) for far-separated sites $i$ and $j$
is therefore ``non-local'' in that density is
transferred from site $j$
to site $i$.
In terms of spin, the state $|i\sigma ,j\sigma' \rangle$ is an 
eigenstate of $S_i^z$ and $S_j^z$ with eigenvalue $\sigma$
and $\sigma'$,
respectively.   The expectation value of spin at other sites is zero.
These are ``spinons.''

Making charged excitations without spin simply requires annihilating
an electron at the center of each spinon.  Anderson's
state with ``holons'' at sites $i$ and $j$ is then
\begin{equation}
|ij\rangle\equiv {\cal N}_{ij} c_{i\sigma} c_{j \sigma'}
|i\sigma,j\sigma'\rangle,
\label{holondef}
\end{equation}
where ${\cal N}_{ij}$ is a normalization factor chosen so that 
$\langle ij|i'j'\rangle = \delta_{ii'}\delta_{jj'}+
\delta_{ij'}\delta_{ji'}$.
Note that the choice of $\sigma$ and $\sigma'$ in
Eq.\ (\ref{holondef}) is arbitrary, 
apart from an overall phase factor.
The holons defined by Eq.\ (\ref{holondef}) are unconventional
quasiparticles with novel statistics
originating from the non-local Gutzwiller projection procedure.

{\it Scheme 2} -- An alternate scheme for constructing spinons and 
holons begins from Slater determinants with uniform charge density
but {\it non-uniform} spin density.\cite{dsr} 
In this approach the average
charge density is one particle per site even before projection, and
the non-locality occurs in the making of the pre-projected Slater
determinant.

Spinons are constructed by considering tight binding models like
${\cal H}_\chi$, but with additional $\pi$ fluxes localized at the
spinon locations.  The resulting Hamiltonian ${\cal H}_d$
is illustrated in fig. \ref{dantbm}, and the energy spectrum with and
without additional fluxes in fig. \ref{chiraldos}. Adding
the localized
fluxes leads to the appearance of midgap states as shown in fig.
\ref{chiraldos} which are localized around the fluxes.
Despite the absence of translational symmetry in ${\cal H}_d$,
the half-filled ground state $|\chi_d\rangle$ has
uniform charge density,\cite{dsr} because of particle-hole symmetry.
(This contrasts with the {\it non}-uniform charge
density in the pre-projected 
spinon state $c_{i-\sigma} c_{j\sigma'}^\dagger |\chi\rangle$
in scheme 1.) The spin density is non-uniform due to the
singly-occupied midgap state, which is localized near the defective 
plaquette.
As in scheme 1, spinons must be constructed in pairs, since the net
flux through a periodic system must be an integer multiple of $2\pi$.
The spinon pair must be connected by a string of negated 
$\chi_{ij}$'s to maintain a uniform field around each
spinon (see fig. \ref{dantbm}).
Essentially, the non-uniform charge density in the scheme 1
pre-projected spinon state (which leads to non-locality in the
subsequent Gutzwiller projection) 
has been replaced in scheme 2 by magnetic flux insertion,
which has long-range consequences for the tight binding
matrix elements.

Adding these localized fluxes is analogous to Laughlin's
procedure\cite{FQHE} 
for creating quasiparticles in a fractional quantum Hall state,
although 
here only {\it half} a flux quantum is added.  
Since Gutzwiller projecting a locally charge-neutral state should not
qualitatively change the long-range correlations of the state,
the extra $\pi$ flux at the defective plaquettes
should survive Gutzwiller 
projection, resulting in localized excitations with
``semion'' statistics 
halfway between fermion and boson.
Charged holons are created in this scheme by removing a site at the 
center of the added localized flux.

\begin{figure}[b]
\setlength{\unitlength}{0.0125in}%
\begin{picture}(200,310)
\end{picture}
\caption{
Schematic of defect Hamiltonian $H_d$
used in scheme 2 (described in the text) for creating spinons and
holons. (a) A localized flux is added to a plaquette, identified here
by a small open circle. To do this, each link between sites crossed by
the Dirac string (dashed line)
originating at the added flux is multiplied by $-1$. At the
other end of the string another localized flux is created (not shown).
(b) A site is deleted from the system by cutting all links to it
(removing all hopping matrix elements to surrounding sites).}
\label{dantbm}
\end{figure}

Despite scheme 2's intuitive appeal, we will not use it further,
since we are interested in off-diagonal matrix elements of the hopping
Hamiltonian Eq.\ (\ref{htj2}) between holon states at different
positions.  In scheme 2, this would require working in a mixed basis
of two Slater determinants that are ground states of different
pre-projected Hamiltonians.
Remarkably, it can be shown for a special value of $m_0$ that
schemes 1 and 2 are essentially equivalent (details are given 
in the appendix).
In this work, all our calculations use scheme 1 
(Eqs. \ref{prespinon}, \ref{spinon}, and \ref{holondef}).

\begin{figure}[t]
\setlength{\unitlength}{1mm}%
\begin{picture}(70,88)
\end{picture}
\caption[cdcap]{
(a) Density of states resulting from the chiral flux Hamiltonian
${\cal H}_\chi$, Eq.\ (\ref{hmf}) when $m_0=1$ (solid line).
The dotted line spectrum was computed on a 512-site system
with periodic
boundary conditions and broadened with a width $0.01$.
(b) Density of states resulting from the same 512-site system but
with a pair of $\pi$ flux defects, ${\cal H}_d$ as illustrated in fig.
\ref{dantbm}.  Fluxes have been placed as shown in the inset.
Adding the fluxes leads to the formation of midgap states
which split off from the upper and lower band, as shown.
See the appendix for further discussion of these midgap states.
}
\label{chiraldos}
\end{figure}

\section{HOLON STATISTICS}
\label{sec_statistics}

The charge motion in a spin liquid is governed by
the hopping Hamiltonian Eq.\ (\ref{htj2}).
The off-diagonal matrix elements of ${\cal H}$ in the holon
variational basis $|ij\rangle$ contain
information regarding the kinetic energy and statistical phase of
holons. In particular, these matrix elements are {\it not} simply
given by the bare $t_{ij}$.
The effective matrix element for one holon hopping from site $i$ to
site $j$ in the presence of another holon fixed at site $k$ is given
by
\begin{equation} 
T_{k\,ij} = -\langle jk|{\cal H}_{\rm hop}|ik\rangle,
\label{thop}
\end{equation}
where $i$ and $j$ are nearest-neighbor sites.

We evaluate Eq.\ (\ref{thop}) using a me\-thod described
previously.\cite{us}
Briefly, we consider a cluster of $N_\Gamma$ sites embedded in an
infinite system.  Correlations within the cluster are
computed exactly,
while correlations outside the cluster are accounted for
only approximately.
The thermodynamic limit is taken by considering clusters of various
shapes and sizes centered on the sites $i$, $j$, and $k$, 
and extrapolating
to the infinite cluster limit.   In practice, this limit is
reached when 
the cluster size exceeds the correlation length $\xi_\chi$ in the 
pre-projected Slater determinant $|\chi\rangle$.  This method is 
substantially more efficient than exact projection of Slater
determinants on finite toroidal systems (see below).

The effective holon hopping matrix element $T_{k\,ij}$ contains 
information about holon statistics.\cite{read} Computation
of $T_{k\,ij}$
allows a direct check of the fractional statistics invoked in
the picture of Zou {\it et al.}.\cite{zou}
The hopping Hamiltonian Eq.\ (\ref{htj2}) restricted to the
holon pair basis Eq.\ (\ref{holondef}) is
\begin{equation}
{\cal H}_T = -\sum_{ijk} T_{k\,ij} |kj\rangle\langle ki|.
\end{equation}
In general, we can separate $T_{k\,ij}$ into a magnitude and a phase,
\begin{equation}
T_{k\,ij} = |T_{k\,ij}|
\exp \left(i\int_i^j{\bf A}(k,{\bf l})\cdot d{\bf l} \right),
\end{equation}
where the phase is interpreted as the line integral of an effective
vector potential, ${\bf A}$.
This vector potential can be separated\cite{arovas} into two distinct
parts: ${\bf A}(k,{\bf l})={\bf A}_0({\bf l})+{\bf A}_S(k,{\bf l})$.
The part of ${\bf A}$ which is independent of the position of site $k$
is absorbed into ${\bf A}_0$, and corresponds to a uniform background
magnetic flux through which the holons move.
The remaining part ${\bf A}_S$ depends on the relative holon position.

Since the holon wavefunctions Eq.\ (\ref{holondef}) can be
made symmetric
under holon interchange,\cite{holonsign} the vector potential
${\bf A}_S$ 
contains complete information about the holon statistics.  
Consider moving one holon around a partner which is held
fixed at site $k$.  
The net phase factor for hopping one holon completely around another
fixed holon is $\oint {\bf A}_S\cdot d{\bf l}\equiv 2\Phi_S.$
When the moving holon path is made infinitely large, $\Phi_S$ 
is the statistical phase for holon interchange.\cite{read} 

To implement this program, we first compute the
background vector potential ${\bf A}_0$ by calculating $T_{k\,ij}$ 
when $k$ (the fixed holon location) is far from $i$ and $j$ 
(the initial and final moving holon locations).
We find numerically that the background flux is indistinguishable from
the fictitious uniform flux introduced in the pre-projected state. 
The vector potential ${\bf A}_0$ corresponds to a uniform flux of
$\pi$ through each elementary plaquette,
as explicitly included in ${\cal H}_{\chi}$.
To compute ${\bf A}_S$ we again calculate\cite{phasefn} $T_{k\,ij}$'s
on a path circling the fixed partner holon,
and subtract the contribution from ${\bf A}_0$.

\begin{figure}
\setlength{\unitlength}{1mm}%
\begin{picture}(70,63)
\end{picture}
\caption{
Statistical field density $\oint_C {\bf A}_S\cdot d{\bf l}$ through
the elementary plaquettes surrounding a fixed holon, when
$m_0=1$ (filled circles), $m_0=0.4$ (open squares), and 
$m_0=0.2$ (open circles).
Monte Carlo statistical error bars are approximately the same size as
the circles.  The inset illustrates the positions
of the plaquettes whose field density has been plotted; each data
point is accompanied by a letter corresponding to its location.}
\label{fluxmap}
\end{figure}

What is the ``form factor'' of the statistical flux bound to each 
holon?  One might expect that the bound flux would be distributed
in a region of linear extent $\xi$ about the holon.  The statistics 
would only be well-defined for holons separated by this distance.
This situation is similar to the case of the $^4{\rm He}$ atom, 
which is a composite of six fermions that behaves as a boson only
on length scales longer than a Bohr radius, so that the internal
atomic structure is essentially fixed.

Surprisingly, we find that for the holon wavefunctions
Eq.(\ref{holondef}) the statistical flux around each holon
is sharply peaked in its immediate vicinity, and that the radius
of the bound flux tube is the lattice spacing, not the spin
correlation
length.
Figure~\ref{fluxmap} shows the statistical field density
$\oint_C {\bf A_S}
\cdot d{\bf l}$ where the path $C$ encloses one plaquette, plotted
against the plaquette's geometrical distance from the origin.
Most of the statistical flux is concentrated in the core
region immediately surrounding the holon (four plaquettes).
Thus the phase shown in fig. \ref{statphase} represents
the total statistical
phase for far-separated holons.

Figure \ref{statphase} shows the statistical flux $\Phi_S$ threading a
small $2\times2$ square loop around a fixed holon as a
function of $m_0$.
For small $m_0$, the spin-spin correlation length tends
to infinity, and
the statistical flux through this small path vanishes.  For $m_0$ of
order unity, the spin-spin correlation length is essentially
the lattice
constant, and the statistical flux nears $\pi/2$.

A slight negative contribution may be present outside the core (see 
fig. \ref{fluxmap}, especially the filled points
representing $m_0=1$).
We are unable to determine whether
this contribution is present
for inter-holon separations exceeding $\sim 5$ lattice constants.
Thus we cannot rule out the possibility that a small statistical
flux density at large distances from a holon integrates to
give a significant contribution to $\Phi_S$.  The existence of 
such a long-range tail to the flux distribution could make the
holon statistics ill-defined, since the net phase would depend
on the inter-holon separation.
Neglecting this possible long-range part of the flux distribution, 
we conclude that the effective statistics of the excitations
defined by Eq.\ (\ref{holondef}) are continuously adjustable
between values of zero (corresponding to a boson) and $\pi/2$ 
(a ``semion'').

Using an analogy with the fractional quantum Hall effect,
Laughlin and Zou have argued\cite{zou} that the holon statistics
in chiral spin liquids should be {\it quantized} as semions.
Although the
logic of this argument seems strong, our calculations (and the 
numerical calculations
of ref. \onlinecite{zou}) find a statistical phase that is 
{\it not} simply $\pi/2$.
Since here we have carefully considered extrapolation to
the thermodynamic
limit, this is not the source of the discrepancy.
There are two possible
resolutions: (1) there is a weak, longer-range tail to the statistical
flux distribution which we cannot resolve in our calculations,
but which
adds up to enforce $\Phi_S = \pi/2$, or (2) the
variational states we have
considered are {\it not} good holon states for small $m_0$,
but only for 
$m_0$ of order unity.  It is only for such $m_0$ that the
correspondence
between scheme 1, scheme 2, and the construction of Laughlin
and Zou\cite{zou}
has been firmly established.

\begin{figure}[t]
\setlength{\unitlength}{1mm}%
\begin{picture}(70,73)
\end{picture}
\caption{
Total phase accumulated by hopping one holon around another for the
shaded loop shown in the inset, as a function of the mass
parameter $m_0$.
This is the core contribution to twice the statistical
phase, $2\Phi_S$.
The filled circles are computed using the cluster method, after
extrapolation to the thermodynamic limit.
The open circles represent the same quantity calculated
on an $8\times 8$ 
torus, which suffers from a discrete sampling effect and
is inaccurate (see text).
Monte Carlo statistical error bars are approximately the same size as
the circles.  The lines are guides to the eye.}
\label{statphase}
\end{figure}

Our embedded cluster results agree with calculations performed
on finite tori when properly extrapolated to the thermodynamic limit.
While the two methods converge to the same result, the
toroidal calculations
converge much more slowly, and with important finite size
effects.  (This is also true for spin-spin correlations.\cite{us})
To confirm this we calculated the core contribution to the statistical
phase for $L\times L$ tori with $L =$ 4, 6, 8, 10, 12, 14, and 16.  
To confront the worst-case scenario for finite size effects we studied
small $m_0$ in detail, where the correlation length $\xi$ is
longest.  
Fig. \ref{extra} shows the statistical phase from the nearest four
plaquettes for $m_0 = 0.01$ and various toroidal systems.
The embedded cluster method rapidly converges
to zero statistical flux, 
as does the the sequence of tori with $L = 4n+2 =$ 6, 10, 14, ...  
The sequence of tori with $L = 4n = $ 4, 8, 12, 16, ... also
extrapolate to zero, but surprisingly slowly.
This is likely due to the discrete Brillouin zone sampling at 
special symmetry points for these tori, a flaw which is not present 
in our embedded cluster method.  This accounts for the
apparent discrepancy
between the 8 $\times$ 8 torus results and the embedded cluster method
in fig. \ref{statphase}.
(The $4\times 4$ torus is a particularly 
extreme case: the Slater determinant ground state wavefunction of
$H_\chi$ is then {\it independent} of $m_0$ due to
this sampling effect.)  From this calculation we conclude that the
finite tori and embedded cluster methods agree when extrapolated to
the infinite system limit.

\begin{figure}[t]
\setlength{\unitlength}{1mm}%
\begin{picture}(70,54)
\end{picture}
\caption[extra-cap]{
Extrapolation to the thermodynamic limit $N\rightarrow\infty$
of the core contribution to the statistical flux
(shown in fig. \ref{statphase}) when $m_0=0.01$. ($N$ is the number of
sites in the projected Slater determinant.) Open squares
reresent $4n\times
4n$ toroidal systems, solid squares $(4n+2)\times (4n+2)$ toroidal
systems.  The $4n\times 4n$
toriodal systems suffer from an effect resulting from the discrete sum
over the single particle Brillouin zone of $H_\chi$, Eq.\ (\ref{hmf}).
Open circles indicate the result obtained with $N$-site
circular clusters
embedded in an infinite system. The dashed line shows the result for a
small embedded cluster ($N=13$), at which point convergence
has already
been reached for the phase information in $T_{k\,ij}$.
The Monte Carlo statistical error bars are smaller than the markers.}
\label{extra}
\end{figure}

\section{PAIR STATES}
\label{sec_pair}

We can now compute the pair wavefunction for two holons
moving through 
a chiral spin liquid background by simply numerically diagonalizing 
${\cal H}_T$ in the two-holon basis Eq.\ (\ref{holondef}).
We find that the lowest-energy two-holon bound state is a complex
$d$-wave state which breaks T and P symmetries.\cite{dsr2} 

To diagonalize ${\cal H}_T$, we note that this two-body Hamiltonian is
invariant under the same set of (magnetic) lattice translations
as $H_\chi$. 
Bloch's theorem then states that each pair wavefunction can be
labeled by a center of mass momentum ${\bf K}$, and must have the form
\begin{equation}
|\Phi \rangle_{\bf K} = \sum_{i<j}
e^{i{\bf K}\cdot ({\bf r}_i + {\bf r}_j)/2} 
\phi(ij) | ij \rangle ,
\label{define_psi}
\end{equation}
where ${\bf r}_i$ is the position of site $i$, and the relative
wavefunction $\phi(ij)$ depends only on the lattice translation
bringing site $j$ into the same magnetic unit cell as site $i$,
as well as the basis indices of the two sites.

By diagonalizing ${\cal H}_T$ we find the doubly charged ground state 
$|\Phi\rangle$ in our variational basis. This calculation of two
holons in a chiral spin liquid resembles in spirit
Cooper's original demonstration that a pairing instability exists
for two electrons attracting above a Fermi sea.
The effective holon hopping Hamiltonian ${\cal H}_T$ provides a 
kinetic resonance energy of order a few percent
$t$ for nearby holons, because
the magnitude $|T_{i\,jk}|$ is enhanced when sites $i$ and $j$ 
(or $i$ and $k$) are close (see table \ref{tstatic}).
Also, we have up until now neglected
a static interaction $V_{ij}$ between holons at sites $i$ and $j$
which arises from the spin interaction, Eq.\ (\ref{htj1}):
\begin{equation}
V_{ij} = J\sum_{\langle lm\rangle}
\big (\langle ij|{\bf S}_l\cdot {\bf S}_m
|ij\rangle - \langle\chi | P_G {\bf S}_l\cdot 
{\bf S}_m P_G |\chi\rangle\big ).
\end{equation}
Roughly speaking, $V_{ij}$ provides a short-range static attraction
because neighboring holons can share a broken
antiferromagnetic bond. In the undoped spin liquid,
each unbroken bond contributes an energy
$J\langle {\bf S}\cdot{\bf S}\rangle_{\rm NN}
=(-0.205 \pm 0.005)J$ when
$m_0=1$.  To estimate the static holon potential we compute
$\langle ij|{\bf S}_l\cdot {\bf S}_m|ij\rangle$ for the
nearest-neighbor sites $(lm)$ around a holon pair fixed
at sites $i$ and $j$ using the embedded cluster method.
Table \ref{tstatic} summarizes the results 
when $m_0=1$.  A far-separated holon pair (5 lattice constants
apart) breaks eight antiferromagnetic bonds, costing
$8\times |J\langle {\bf S}\cdot {\bf S}\rangle_{\rm NN}|\sim 1.6 J$,
but the remaining nearby bonds are strengthened, so
isolated holons end up costing slightly less energy to create.
As the holons are brought closer together, they eventually
share a broken bond, causing a net attractive potential of strength
$\sim 0.3 J$.
Since both the kinetic resonance and the static
potential are attractive, the variational ground state
$|\Phi\rangle_{\bf K}$ is a bound state of two holons.
Numerically, we find that the lowest energy bound state has
center of mass momentum ${\bf K}=0$, and 
effective mass $m^* \equiv 1/(\partial^2 E/\partial K^2)\sim 3m$,
where $m=4|T|$ is the effective mass of an isolated holon with
hopping matrix element $T$, evaluated for far-separated holons.

\begin{table}
\caption{Holon interactions. Hopping matrix element $T$ for one holon
fixed at site $i$ and another moving from site $j$ to site $k$ when
$m_0=1$. In the radial case sites $i$, $j$, and $k$ are collinear,
in the azimuthal
case the hopping direction $(jk)$ is perpendicular to $(ij)$. In each
case $d_{jk}>d_{ij}$.}
\begin{tabular}{cccc}
 & \multicolumn{2}{c}{$|T_{i\,jk}|/t$} \\
$d_{ij}$ & radial & azimuthal & $V_{ij}/J$ \\
\hline
1 & $0.814\pm 0.003$ & $0.827\pm 0.004$ & $1.20\pm 0.08$ \\
2 & $0.806\pm 0.002$ & $0.804\pm 0.004$ \\
3 &                  & $0.811\pm 0.002$ & $1.36\pm 0.1$ \\
5 & $0.808\pm 0.002$ & $0.805\pm 0.003$ & $1.45\pm 0.1$ \\
\end{tabular}
\label{tstatic}
\end{table}

\begin{figure}[t]
\setlength{\unitlength}{1mm}%
\begin{picture}(70,132)
\end{picture}
\caption[overlap-cap]{
Overlap between $|(ij)\rangle$ (a singlet hole pair,
Eq.(\ref{holedef}))
and $|ij\rangle$, (a holon pair, Eq.(\ref{holondef}))
as a function of geometrical distance $d_{ij}$
between sites $i$ and $j$ when (a) $m_0=1$, (b) $m_0=0.4$,
(c) $m_0=0.2$.  For a given pair, site $j$ may lie on the
same sublattice 
(open circles) or the opposite sublattice (filled circles) as
site $i$.
The correlation length $\xi_\chi$ is indicated in each frame.
The dominant overlap occurs for nearest-neighbor pairs, and is close
to unity over the entire range of $m_0$.  The overlaps are large
for pairs closer than $\xi_\chi$, and are slightly larger
for opposite sublattice pairs than for same sublattice pairs.
The Monte Carlo statistical error bars are approximately the same size
as the circles.}
\label{overlap}
\end{figure}

What is the symmetry of the pair state $|\Phi\rangle$?  To answer
this question, consider the relative pair wavefunction $\phi(ij)$.
Since the relative phases of the holon pair states $|ij\rangle$
for different $ij$ depend on the arbitrary choice of gauge used
to construct the Slater determinant $|\chi\rangle$, the relative
phase of $\phi(ij)$ are also gauge dependent.  This complication
can be eliminated by projecting the two-holon state $|\Phi\rangle$
into the set of states comprising a singlet {\it hole} pair
added to the undoped spin liquid:
\begin{equation}
|(ij)\rangle = {\cal M}_{ij}\left ( 
  c_{i\uparrow}c_{j\downarrow} - 
  c_{i\downarrow}c_{j\uparrow} \right )P_{\rm G} |\chi\rangle .
\label{holedef}
\end{equation}
Here ${\cal M}_{ij}$ is a normalization factor chosen so that
$\langle(ij)|(i'j')\rangle = \delta_{ii'}\delta_{jj'}
+\delta_{ij'}\delta_{ji'}$. 
These hole pair states are independent of the gauge choice in 
$|\chi\rangle,$ since $P_G|\chi\rangle$ has precisely one particle
per site, and is independent of gauge choice.  The inner product
\begin{equation}
\Phi(ij) = \langle (ij)| \Phi\rangle
\label{inner}
\end{equation}
is then a gauge invariant description of that part of the
two-holon state which resembles a singlet hole pair.

How similar are two holons and a singlet hole pair?  
The overlap between these two states is simply an expectation
value in the state $|\chi\rangle$
\begin{equation}
\langle (ij)|ij\rangle = 
2 {\cal N}_{ij} {\cal M}_{ij}
\langle\chi| P_{\rm G} (1-n_{i\uparrow})n_{j\uparrow} 
c^\dagger_{i\downarrow} c_{j\downarrow}|\chi\rangle,
\label{overlap_def}
\end{equation}
and can be computed using the embedded cluster me\-thod:
The magnitude of this overlap versus inter-holon distance $d_{ij}$ 
is plotted in fig. \ref{overlap} for three different values of $m_0$.
A nearest-neighbor holon pair is essentially identical to a
singlet hole pair, as shown by the near-unit overlap in
fig. \ref{overlap} at $d_{ij}=1$ in all three cases.
The overlap decays quickly as 
sites $i$ and $j$ are separated, presumably since far-separated 
holons disrupt the system differently than holes.  Holes maintain 
a strong spacial correlation between spin and charge, while the holon
has no spin.  When holons are close together, however, the composite
object has charge $2e$ and spin-0, permitting an overlap with a
singlet hole pair with the same quantum numbers.
The decay of the overlap with distance occurs on a length scale 
roughly equal to the correlation length $\xi_\chi$, as shown in fig. 
\ref{overlap}.

\begin{figure}[t]
\setlength{\unitlength}{1mm}%
\begin{picture}(70,80)
\end{picture}
\caption[psym-cap]{
The relative pair wavefunction $\Phi(i,j).$  all $\phi_n$ are real
numbers.  The symmetry is a combination of $d_{x^2 - y^2}$
and $d_{xy}$.}
\label{pairsym}
\end{figure}

Using Eq.\ (\ref{overlap_def}) and the two-holon state $|\Phi\rangle$
obtained by exact diagonalization of ${\cal H}_T$, we can compute
the ``hole pair wavefunction'' (the inner product Eq.\ (\ref{inner})) 
for inter-holon separations up to $4$ lattice constants.
Figure \ref{pairwave} shows the radial probability density
$|\Phi(ij)|^2$ as a function of the inter-holon distance $d_{ij}$.
The amplitude is mostly concentrated near the origin, due
to the kinetic resonance effect already mentioned and the short range
of the holon pair/hole pair overlap.
Further static local attraction due to the sharing of
broken bonds only enhances this effect, without changing the
angular symmetry of the state, as we verified numerically
by considering additional short-range static attraction mimicking
$V_{ij}$.

\begin{figure}[t]
\setlength{\unitlength}{1mm}%
\begin{picture}(70,63)
\end{picture}
\caption[pwave-cap]{
Probability density in the two-holon wavefunction $\Phi(ij)$
versus inter-holon separation for holons governed by
${\cal H}_T$ with $m_0=1$ (filled circles), $m_0=0.2$ (open circles). 
In each case the wavefunction's angular symmetry is $d$-wave as 
illustrated in fig. \ref{pairsym}.}
\label{pairwave}
\end{figure}

For all values of $m_0$, the relative hole pair wavefunction
$\Phi(ij)$ 
has $d$-wave symmetry (i.e., the state changes sign under 90 degree
rotation), as illustrated in fig. \ref{pairsym}.
Because of the explicit breaking of time-reversal and parity in 
the chiral spin liquid, this state is a complex linear combination
of $d_{x^2-y^2}$ and $d_{xy}$ states.  (Although these states 
transform as different irreducible representations of the tetragonal
group, they are mixed in the group of four-fold rotations that 
remains when reflections are removed from the tetragonal
group.\cite{dsr2})
As shown in fig. \ref{pairsym}, the relative pair wavefunction 
is purely real on the $x$ and $y$ axes, and purely imaginary along
the diagonals; $\Phi(ij)$ is permitted by symmetry to be a general
complex number elsewhere, but we find that it remains either purely
real or imaginary in non-symmetry directions as well.

Since the two-holon state $|\Phi\rangle$
has substantial overlap with a two-hole
state, it is natural to consider a multi-holon state obtained by 
macroscopically occupying the hole pair wavefunction $\Phi(ij)$, 
thus arriving at a BCS superconductor with a $d$-wave Cooper
wavefunction.
To show that this is the ground state
of ${\cal H}$ or a similar spin Hamiltonian requires a much more 
elaborate calculation than we have given here.  Nevertheless, if we
assume that the ``anyon condensate'' of Laughlin {\it et al.} is
obtained by a Bose condensation of holon pairs, then we may infer
from our calculation that such a state will have {\it conventional}
electron-pair off-diagonal long-range order:
\begin{equation}
\lim_{|(ij)-(i'j')| \rightarrow \infty}
\langle c_{i\uparrow}^\dagger c_{j\downarrow}^\dagger
c_{i'\downarrow} c_{j'\uparrow} \rangle \rightarrow \Psi^*(ij)
\Psi(i'j').
\label{ODLRO}
\end{equation}
The order parameter $\Psi(ij)$ is nothing more than the internal
Cooper pair wavefunction, which for dilute, non-overlapping
pairs should be given simply by the relative hole-pair wavefunction
$\Phi(ij)$ we have calculated above.  Thus a holon condensate is
expected to possess all of the features of a BCS superconductor
that has the unconventional $d$-wave order parameter described above.
Most strikingly, such a $d$-wave state is expected to have a full
energy gap, since there are no nodes in the Cooper pair wavefunction.

\section{CONCLUSIONS}
\label{sec_conclude}

We have investigated the properties of charged excitations in a 
chiral spin liquid on a square lattice, using
a variational basis which explicitly separates the spin and
charge degrees of freedom. By directly computing the off-diagonal 
Hamiltonian matrix elements of the holon excitations, 
we found that the statistics of the excitations described by
these variational states vary {\it continuously} between boson 
($\Phi_S=0$) and semion ($\Phi_S=\pi/2$).  
The gauge field responsible for the statistical phase is localized in 
the area surrounding the holon, with a core size comparable to the
lattice spacing.

These results may be intuitively understood by viewing holons
and spinons
as topological excitations in a chiral tight-binding model.\cite{dsr} 
In the appendix we show that this picture (while computationally 
more difficult to implement) is an essentially equivalent
representation 
of Andersons holons for spin liquids with short spin-spin correlation 
lengths.

The two-holon ground state has total momentum zero, and its
relative wavefunction transforms as a linear combination of the
$d_{x^2-y^2}$ and $d_{xy}$ representations of the tetragonal group.
Such a linear combination is permitted\cite{dsr2} by the explicit
breaking of T and P symmetry in the chiral spin liquid.
Two factors contribute to the formation of a two-holon bound state: 
(a) a static attraction of order $J$ between holons,
and (b) a kinetic resonance due to enhanced holon
hopping in the vicinity of another holon.\cite{marsiglio}
These effects are reminiscent of a ``spin bag,'' in that the local 
environment around
one holon lowers the energy of a second in its vicinity.

The pair wavefunction of a bound holon pair has significant
overlap with a bound singlet hole pair, especially at short
distances.  This overlap suggests that a holon pair condensate
is closely related to a hole pair (or electron) condensate, as
found in the BCS theory of superconductivity.  We conclude that
a superconductor arising from a pair condensate of bound holons 
is a BCS-like superconductor with a have a gap function with $d$-wave 
symmetry.\cite{dsr2}
Such a superconductor would be fully gapped, since by mixing
to $id_{xy}$ the gap function avoids crossing zero.
This superconductor
can be described {\it either} as a $d$-wave BCS superconductor or as
an anyonic superconductor. These two languages are evidently 
equivalent ways to describe the same state.

\acknowledgements
We thank K.G.~Singh for useful discussions, and P.E.~Lammert
for useful
discussions and a careful reading of the manuscript.
Work at Berkeley was supported by the NSF under grant DMR-91-57414 and
by the US Dept. of Education.
D.S.R. acknowledges a grant from the Alfred P. Sloan Foundation.

\appendix
\section*{EQUIVALENCE OF HOLON SCHEMES}

In this appendix we show that 
the Anderson-holon  and the topological charged excitations
described previously in ref. \onlinecite{dsr} are equivalent. 
This equivalence provides an intuitive
picture of the Anderson-holon excitation and 
some insight into its local structure.
The argument relies heavily on the equivalence between an
Anderson-holon
wavefunction and the Kalmeyer-Laughlin wavefunction established
by Laughlin and Zou\cite{zou} so we start by briefly recalling the 
content of their proof. 

In ref.\onlinecite{zou} it is shown that the state obtained by
Gutzwiller projecting the ground state of
$H_\chi$ at half-filling and with
$m_0=2 e^{-\pi /4}\approx 0.912$
is identical to the Kalmeyer-Laughlin (KL) state.\cite{chifal}
In first quantization, the KL state is expressed by the wavefunction
\goodbreak
\begin{eqnarray}
\psi _{KL}(z_1, \dots , z_{N/2}) &=&\nonumber\\
\prod_{ \mu < \nu} ^{N/2} 
    (z_\mu - &z_\nu&)^2 \prod_{\gamma = 1}^{N/2}
    G(z_\gamma) \exp \left[ - { |z_\gamma  |^2 \over 4 } \right],
\label{a1}
\end{eqnarray}
where $z_\gamma = (m_\gamma + i n_\gamma) $ is a complex number
defining the position of an up spin labeled by $\gamma$ at square
lattice coordinates $(m,n)$, 
and $G(z)$ is a gauge-dependent
function. $G(z)=(-1)^{(n+m+nm)}$ in the gauge chosen by the authors of
ref.\onlinecite{zou}.

The proof consists of two steps. In the first it is shown that
given the first-quantized version of the {\it up spin} part of the
half-filled ground
state of $H_\chi$ -- a function  $\phi(\{{\bf r}\})$ where the set of
coordinates
$\{{\bf r}\} \equiv {\bf r}_{1},\dots , {\bf r}_{N/2}$ 
describes the position of $N/2$ up
electrons on a lattice of $N$ sites -- one can write down 
the wavefunction corresponding to the Gutzwiller projected state
$\Psi$ as a
function of the same set of coordinates:
\begin{equation}
\Psi( \{ {\bf r} \} ) = \phi ^2 ( \{ {\bf r} \} )
e^{i f( \{ {\bf r} \} )},
\label{a2}
\end{equation}
where the function $f(\{{\bf r}\} )$ depends on the choice of
the basis used
to represent $\Psi$. (See Eq.\ (\ref{a12}) for a precise definition.)
In the second step it is shown that with a particular
choice\cite{chifal} of 
$m_0$, the Slater determinant obtained by filling the lower
band of $H_\chi$ coincides with the determinant
obtained by filling the energy levels corresponding to an electron
in free space moving in a uniform magnetic field with
$l=(\hbar c /e B)^{1/2} =1/\sqrt\pi $, provided the single-particle 
wavefunctions defined on a continuum are restricted to points on a 
square lattice with lattice constant equal to unity.
This representation of $\phi$, together with Eq.\ (\ref{a2}) is then
used to  show that $ \Psi=\psi_{KL}$.
As a simple corollary, Zou and Laughlin show that
the wavefunction corresponding to an Anderson-holon pair at sites
$A$ and $B$ is represented by
\begin{eqnarray}
|z_A,z_B \rangle =\prod_\gamma (&z_\gamma& - z_A )(z_\gamma - z_B)
      \prod_{ \mu < \nu} (z_\mu - z_\nu )^2\nonumber\\
  &\times& \prod_{\gamma }
    G(z_\gamma) \exp \left[ -{|z_\gamma |^2 \over 4 } \right].
\label{a3}
\end{eqnarray}

Our approach is very similar. First we show that the first
step follows generally when a particle-hole symmetry operator 
exists, thus it also applies to the
class of Hamiltonians $H_d$ described in ref.\onlinecite{dsr}.
Second, we argue that a representation of the filled lower
band of $H_d$ is obtained by filling the first Landau
level corresponding to an electron in a uniform field with two
flux tubes of appropriate strength at position ${\bf r}_A$
and ${\bf r}_B$,
again having restricted the wavefunction to points on a square
lattice. Although we will not prove this point rigorously it is
a natural assumption in view of the results of ref.\onlinecite{zou}.

The Hamiltonian $H_d$ describing the topological charged excitation
of section \ref{sec_general}'s scheme 2 
is obtained from $H_\chi$ by adding flux defects in the form of
an additional half-flux quantum $\pi$
through a plaquette adjacent to site $A$, the same through a plaquette
adjacent to site $B$, and a string of $\pi$ phase additions to the
phases of the $\chi_{ij}$'s between sites $A$ and $B$
to restore the proper flux in the remaining plaquettes.\cite{dsr}
(The actual position of this string is irrelevant, as it can be moved
by a suitable gauge transformation.) Note that the added
fictitious flux couples only to the particles' orbital motion.

It is convenient to start with $H_\chi$ (and 
$H_d$) expressed in a gauge such that the $\chi_{ij}$ are
real for nearest-neighbor and imaginary for 
next-nearest neighbor bonds.
This choice of phases for the link variables allows a simple
definition 
of the particle-hole conjugation operator $\Upsilon$,
which we define to be the {\it linear} operator satisfying
\begin{eqnarray}
\Upsilon |{\rm vac}_\sigma \rangle &=& e^{i\alpha}
|F_\sigma\rangle, \nonumber\\
\Upsilon ^{-1} c_{{\bf r }\sigma}\Upsilon &=& e^{i{\bf r} \cdot
{\bf Q}} c_{{\bf r }\sigma}^\dagger ,
\label{a4}
\end{eqnarray}
where $\alpha$ is an arbitrary phase, ${\bf Q} \equiv (\pi,\pi)$,
$|{\rm vac}_\sigma \rangle$ is the vacuum state in the $\sigma$ spin
subspace, and 
$|F_\sigma\rangle = \prod _{ \{{\bf r}\} }c_{\bf r}^\dagger 
|{\rm vac}_\sigma \rangle $ is the state with one $\sigma$
electron at each
of the $N$ sites in the system.
To avoid ordering ambiguities this product is performed
in an (arbitrary) order specified by an ordering function $O({\bf r})$
numbering the sites of the lattice from $0$ to $N-1$ such that
$ O({\bf r}_1 ) < O({\bf r}_2 )< \cdots < O( {\bf r}_N ) $.

With this definition of $\Upsilon$ one can show that $\Upsilon$ is
unitary and $\Upsilon ^2 =e^{i\gamma }$, where $\gamma$ depends only
on $N$ and $\alpha$.
The usefulness of $\Upsilon$ is apparent when we realize
that $\Upsilon$
is a symmetry operation for both ${\cal H}_\chi$ and ${\cal H}_d$:
$[ \Upsilon , H_\chi ] = [ \Upsilon , H_d ] = 0$.
Notice that both charge 
density and current density are odd under particle-hole symmetry, that
is $\Upsilon ^\dagger (1 - n_{\bf r}) \Upsilon = n_{\bf r} -1 $,
$\Upsilon ^\dagger J_{\bf a} \Upsilon = -J_{\bf a}$,
where $J_{\bf a}= i\sum_\sigma ( e^{i\phi_{\bf a}} 
c_{ {\bf r}, \sigma}^\dagger c_{ {\bf r}+{\bf a}, \sigma} - 
{\rm H.c.} ) $.

We also define the {\it antilinear} operator $Q$ :
\begin{eqnarray}
Q |{\rm vac}_\sigma \rangle &=&  |F_\sigma \rangle  , \nonumber\\
Q ^{-1} c_{{\bf r}\sigma} Q &=&  c_{{\bf r}\sigma}^\dagger .
\label{a5}
\end{eqnarray}
$Q$ is antiunitary, anticommutes with $H_\chi$ and $H_d$,
and commutes with $\Upsilon$.
A simple consequence of the commutation relations of $Q$, $\Upsilon$,
and the Hamiltonians $H_d$ or $H_\chi$ is that 
single-particle states always come in pairs, $|E_\sigma\rangle$
and $ Q \Upsilon |E_\sigma\rangle $, corresponding to energies
$E$ and $-E$ respectively. 
Since in the absence of flux tubes and at half filling the
ground state of $H_\chi$
is fully gapped, symmetry suggests (and calculations have explicitly 
shown\cite{dsr}) that with a single flux-tube and an
odd number of sites the energy spectrum of $H_d$
will have a midgap state, whose amplitude is localized near the flux
tube.  With an even number of sites and 
two flux-tubes at ${\bf r}_A$ and ${\bf r}_B$ 
the degeneracy of the two midgap states in the energy spectrum
of $H_d$ is lifted by
$\Delta E _ {\rm loc} $, which is exponentially small 
in $|{\bf r}_A-{\bf r}_B|$. (See fig. \ref{chiraldos}.)
 
Our task now is to evaluate the Gutzwiller-projected two-holon
wavefunction 
of ref.\onlinecite{dsr}. This is obtained by first considering
the ground state at half-filling
$|\chi_d\rangle$ of $H_d$ when fluxes have been added at sites
$A$ and $B$, and these same sites have been removed from the lattice.
Notice that the removal of the two sites eliminates the two
midgap states
and that the ground state is non-degenerate and hence invariant under
$\Upsilon$.

After Gutzwiller projection every 
configuration has a single electron at each site and an
equal number of
up and down electrons. A convenient basis in which to represent
$ P_G |\chi_d\rangle $ is
\begin{equation}
| \{{\bf r}\} \rangle = S_{\tilde{\bf r}_1}^-
\cdots S_{\tilde{\bf r}_M}^- |F_\uparrow\rangle,
\label{a6}
\end{equation}
where $S_{\bf r}^-= c_{{\bf r}\downarrow}^\dagger c_{{\bf r}\uparrow}$
is the usual spin lowering operator,
$\{ {\bf r} \} $ denotes the set of positions of the up spin
electrons, and $ \{\tilde{\bf r}\}$ denotes the set complementary to
$\{{\bf r}\}$ specifying the positions of the down spin electrons.
The sets $\{{\bf r}\}$ and $\{\tilde{\bf r}\}$ each contain
$M= N/2 -1 $ members. 

The Gutzwiller-projected wavefunction in this basis is given by
\begin{equation}
\Psi ( \{ {\bf r} \} )
\equiv \langle \{ {\bf r} \} | P_G | \chi_d \rangle
= \langle \{ {\bf r} \} | \chi_d \rangle,
\label{a7}
\end{equation}
since none of the states $|\{{\bf r}\}\rangle$ contain site
double-occupancy.
Using the anticommutation properties of fermion operators, we find
\begin{equation}
\langle \{ {\bf r} \} |\chi_d \rangle =
\exp \left[ i \pi \sum_{i=1}^M O( \tilde{\bf r}_i ) \right] 
\phi_d ( \{ {\bf r} \} ) \phi_d ( \{ { \tilde {\bf r} } \} ),
\label{a8}
\end{equation}
where
\begin{equation}
\phi_d ( \{ {\bf r} \} ) \equiv 
\langle {\rm vac}_\uparrow | c_{ {\bf r}_M} \cdots
c_{ {\bf r}_1} | \chi_{d\uparrow} \rangle ,
\label{a9}
\end{equation}
and we have made the decomposition
$ |\chi_d\rangle = |\chi_{d\uparrow} \rangle 
\otimes |\chi_{d\downarrow} \rangle$ and dropped the explicit spin
indices on $c_{\bf r}$ and $c^\dagger_{\bf r}$.
We can now use the properties of $\Upsilon$ to obtain
\FL\begin{eqnarray}
\phi_d ( \{ { \tilde {\bf r} } \} ) &=& 
\langle {\rm vac}_\uparrow | c_{ { \tilde {\bf r} }_M } 
\cdots c_{ { \tilde {\bf r} }_1} | \chi_{d\uparrow} \rangle
\nonumber\\
&=& \langle {\rm vac}_\uparrow |
\Upsilon \Upsilon^\dagger c_{ { \tilde {\bf r} }_M } \Upsilon
\cdots \Upsilon^\dagger c_{ { \tilde {\bf r} }_1 }
\Upsilon \Upsilon^\dagger 
| \chi_{d\uparrow} \rangle \nonumber\\
&=&  \exp \left[ i \sum_{i=1} ^M  {\bf Q} \cdot
{ \tilde {\bf r}}_i \right]
\langle F_\uparrow | c_{ { \tilde {\bf r} }_M }^\dagger \cdots 
c_{ { \tilde {\bf r} }_1 }^\dagger | \chi_{d\uparrow} \rangle\\
&=& \exp \left[ i \sum_{i=1}^M {\bf Q}\cdot\tilde{\bf r}_i +
\pi O(\tilde{\bf r}_i) \right ]
\langle {\rm vac}_\uparrow | c_{ {\bf r}_M } \cdots c_{ {\bf r}_1}
  | \chi_{d\uparrow} \rangle,
\nonumber
\end{eqnarray}
which, together with Eq.\ (\ref{a8}), gives
\begin{equation}
\Psi ( \{ {\bf r} \} ) =\exp \left[ -i \sum_{i=1} ^M  {\bf Q} \cdot 
{\bf r}_i \right] \phi_d^2 ( \{ {\bf r} \} ) .
\label{a11}
\end{equation}
In Eq.\ (\ref{a11}) we
have used the freedom in the choice of $\alpha$ in
Eq.\ (\ref{a4}) to eliminate an additional phase independent
of $\{ {\bf r} \}$.

Eq.\ (\ref{a11}) gives the wavefunction $\Psi$ in the particular
gauge in which the link variables $ \chi_{ij} $ are either real or 
purely imaginary. Under any local gauge transformation
$c_{{\bf r}\sigma} \to \exp[-i\Lambda({\bf r})]c_{{\bf r}\sigma}$
we finally derive, with the help of Eq.\ (\ref{a9}), 
\FL\begin{equation}
\Psi ( \{ {\bf r} \} ) = \exp \left[ -i \sum_1 ^M 
{\bf Q}\cdot{\bf r}_i +2\Lambda({\bf r}_i)\right ]
\phi_d^2 ( \{ {\bf r} \};\Lambda ) ,
\label{a12}
\end{equation}
where we have added explicit dependence on
$\Lambda$ to $\phi_d$ in order to remind the reader 
that the up spin part of the ground state
$| \chi_{d\uparrow} \rangle $ is gauge dependent. Eq.\ (\ref{a12}) now
corresponds directly to Eq.\ (\ref{a2}), as desired.

Next, we want to find an alternate representation of $\phi_d$
based on a continuum picture of electrons moving in a uniform
magnetic field with two flux tubes at positions $A$ and $B$.
Notice that although by its definition Eq.\ (\ref{a7}) $\Psi$ is gauge 
invariant, $\phi_d$ is not, and we must be careful to choose a gauge
in the continuum which corresponds to the particular gauge
used to derive
Eq.\ (\ref{a11}) and Eq.\ (\ref{a12}). In the following we
work in the symmetric gauge although the gauge used to derive
Eq.\ (\ref{a11}) and used in
ref.\onlinecite{zou} to express $| z_A , z_B 
\rangle $ is the Landau gauge. (Strictly speaking, this gauge differs
from the Landau gauge by a ``singular'' transformation,
which will be discussed later.)

We start with the continuum Hamiltonian 
\begin{equation}
H = {\hbar^2 \over 2ml^2 } \left( -i\nabla + {\bf A} \right)^2
\label{a13}
\end{equation}
where $ {\bf A } = ({\bf r} \times  {\hat {\bf z}} )/2 + 
{\bf A }_{{\bf r}_A} +{\bf A}_{{\bf r}_B}$,
the unit of length $l=( \hbar c /eB )^{1/2} $, and $B$ has been chosen
so that in the absence of flux tubes the degeneracy of the Landau
levels is equal to $N/2$; this implies $B= \phi_0 /2 $ where
$\phi_0 = hc / e $ is the flux quantum.
The vector potential corresponding
to a flux tube of strength $\alpha \phi_0$ is then given by 
\begin{equation}
{\bf A}_{{\bf r}_A}^\theta = \alpha { \theta_A \over
|{\bf r } - {\bf r }_A |}
\label{a14}
\end{equation}
where $\theta_A$ is the azimuthal angle  measured with
respect to ${\bf r}_A$,
and positive values of $\alpha$ correspond to a localized flux which
strengthens the background magnetic field.

It is convenient to gauge away ${\bf A}_{{\bf r}_A}$ and 
${\bf A}_{{\bf r}_B}$ by 
looking for a solution in the form $\psi ={\tilde\psi}
\exp [ -i \int ({\bf A}_{{\bf r}_A}+ {\bf A}_{{\bf r}_B})
\cdot d{\bf l} ] $.
Then ${\tilde \psi}$ obeys the Schr\"odinger equation for a particle
in an uniform field but must be multivalued in order to make $\psi$ 
single-valued: when encircling ${\bf r }_A$ and ${\bf r }_B$ the phase
of ${\tilde \psi}$ must change by $\alpha_A$ and $\alpha_B$
respectively,
in order to compensate for the phase change due to the
exponential term
in Eq.\ (\ref{a15}). But the most general solution corresponding to 
the lowest Landau level orbitals in the symmetric gauge can be written
as ${\tilde \psi}(z)={\tilde g}(z)e^{z {\bar z}/8} $, where
$z=x+iy$ and
${\tilde g}(z)$ is an arbitrary function of $z$. The solution
of interest is then given by ${\tilde g}(z) = (z-z_A)^{\alpha_A}
(z-z_B)^{\alpha_B}g(z)$, with $g(z)$ analytic everywhere.

Although the solution so defined is valid for any\cite{zeroes} 
value of $\alpha_A$ and $\alpha_B$, our purposes lead us
to consider the special case $\alpha_A = - \alpha_B = \alpha >0 $.
In this case, as $\alpha$ is increased from zero, a lone
state splits off from each of the $n=0$ and $n=1$ Landau
level manifolds:
the first lone state's energy increases, and the second lone state's
energy decreases. 
Note that the degenerate Landau level manifolds can be 
associated with the lower ($n=0$) and upper ($n=1$) band of
the lattice
Hamiltonian $H_d$.  The two lone states which split off are similarly
associated with the two localized states of $H_d$ in the vicinity of
the flux defects.
Filling the $n=0$ Landau level corresponds to computing the Slater 
determinant obtained by filling the levels
\begin{equation}
{\tilde \psi}_k (z)=(z-z_A)^{\alpha}(z-z_B)^{1-\alpha}
z^k e^{z {\bar z}/8}
\label{a15}
\end{equation}
where $k=0,1,\cdots$.
The resulting Vandermonde determinant can be easily
calculated giving the
result
\begin{eqnarray}
\phi(z_1, \cdots , z_M )=
\prod_\gamma (&z_\gamma& - z_A )^\alpha (z_\gamma - z_B)^{1- \alpha }
\prod_{ \mu < \nu} (z_\mu - z_\nu )\nonumber\\
  &\times& \prod_{\gamma }
    G(z_\gamma) \exp \left[ -{|z_\gamma |^2 \over 8 } \right].
\label{a16}
\end{eqnarray}

To complete the argument we now employ the gauge freedom 
associated with $\Lambda$ in
Eq.\ (\ref{a12}). First notice that the use of ${\tilde \psi}$ 
in place of
$\psi$ to obtain Eq.\ (\ref{a16}) is justified because the gauge we 
were effectively using in writing a lattice Hamiltonian with a string
of defects connecting sites $A$ and $B$ is precisely the
singular gauge 
(singular only in the continuum limit, of course) we employed to go
from $\psi$ to ${\tilde \psi}$.
Alternatively we could have used $\psi$ to define $\phi$ in
Eq.\ (\ref{a16})
and performed a gauge transformation on the lattice Hamiltonian
to smooth
out the string connecting $A$ and $B$. When this gauge transformation
is properly taken into account as prescribed by
Eq.\ (\ref{a12}) the result
for $\Psi$ is identical. Finally we have to convert the Landau gauge,
corresponding to Eq.\ (\ref{a11}), into the symmetric gauge
used to express 
$\phi$ in Eq.\ (\ref{a16}). This is easily accomplished
through the gauge
transformation defined by $\Lambda ( {\bf r} )=\pi xy/2$
which, when restricted to the lattice ${\bf r} =(m,n)$, 
makes the prefactor
in Eq.\ (\ref{a12}) equal to 
\FL\begin{equation}
\exp \left[ -i \sum_1 ^M  {\bf Q} \cdot {\bf r}_i
+2 \Lambda ( {\bf r}_i )\right] =
\prod_\gamma (-1)^{m_\gamma +n_\gamma
+m_\gamma n_\gamma}.
\end{equation}
This is identical to the factor $\prod_\gamma
G(z_\gamma) $ appearing in Eq.\ (\ref{a3}).

By using Eq.\ (\ref{a12}) and Eq.\ (\ref{a16}) with the
continuum variables 
$z_\gamma$ restricted
to lattice points and $\alpha = 1/2$, the equivalence between
the two-holon state defined as in Section II, scheme 2, and
the Anderson
holon state, scheme 1, defined by Eq.\ (\ref{a3}) is now
explicitly shown.


\begin{references}
\bibitem{holonsign} A phase factor lurking
in Eq.\ (\ref{spinon}) must be removed by a trivial
phase change to arrive at Eq.\ (\ref{holondef}) with no intrinsic
sign.  This phase factor is $g_{ij}/|g_{ij}|$ for the state
Eq.\ (\ref{spinon}) with spinons at
sites $i$ and $j$, where
$g_{ij}=\langle\chi|c^\dagger_i c_j|\chi\rangle$ is the
lattice Green's function in the underlying Slater determinant.
\bibitem{phasefn} Note that the phase
of $T_{k\,ij}$ generally has smaller Monte Carlo error bars than the
amplitude.
\bibitem{chifal} To be more precise the modulus of the link variables 
$\chi_{ij}$ should not be strictly equal to zero beyond the
next-nearest neighbor distance, but should rather decay to zero 
exponentially in distance. See ref.\onlinecite{zou}.
\bibitem{zeroes} The solutions have to vanish at $z_A$ and $z_B$ but
this can be taken care of by the function $g(z)$.
\end{references}
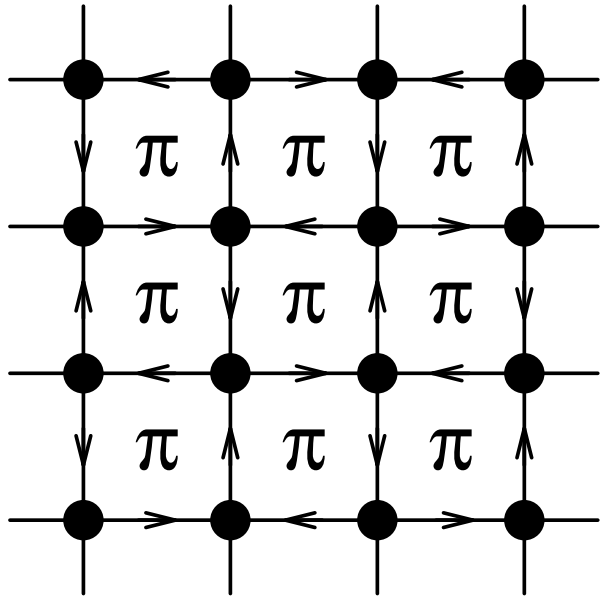
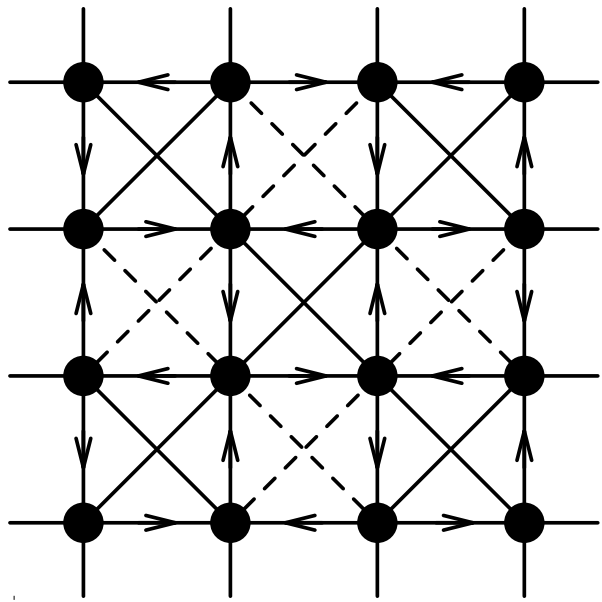
\end{document}